\documentclass[aps,showpacs,floats,preprintnumbers,nofootinbib]{revtex4}
\usepackage{epsfig}

\def\rts{\sqrt s}
\def\lsim{\mathrel{\raise.3ex\hbox{$<$\kern-.75em\lower1ex\hbox{$\sim$}}}}
\def\gsim{\mathrel{\raise.3ex\hbox{$>$\kern-.75em\lower1ex\hbox{$\sim$}}}}

\newcommand{\nc}{\newcommand}
\nc{\baa}{\begin{array}}     
\nc{\eaa}{\end{array}}
\nc{\bew}{\beta_W}
\nc{\dtz}{d^2_{0,0}}
\nc{\dzz}{d^0_{0,0}}
\nc{\dto}{d^2_{1,0}}

\def\mhmax{\mh^{\rm max}}
\def\tev{~{\rm TeV}}
\def\gev{~{\rm GeV}}

\def\bea{\begin{eqnarray}}
\def\eea{\end{eqnarray}}
\def\non{\nonumber}
\def\lsp{\;\;\;\;\;}
\def\beq{\begin{equation}}
\def\eeq{\end{equation}}

\def\re{{\rm Re}~}

\def\wlwl{\wl^+\wl^-\to \wl^+\wl^-}

\def\bce{\begin{center}}
\def\ece{\end{center}}

\def\mplv{M_{Pl\,5}}

\def\phio{\phi_0}

\def\anti{\overline}

\def\mh{m_{h}}
\def\mphi{m_\phi}
\def\what{\widehat}
\def\lw{\Lambda_W}

\def\lphi{\Lambda_\phi}
\def\lbar{\anti\Lambda}

\def\mphi{m_\phi}

\def\hbar{\overline h}

\def\mpl{M_{Pl}}

\def\wl{W_L}

\def\lnda{\Lambda_{\rm NDA}}
\def\call{{\cal L}}

\def\mpl{m_{Pl}}
\def\mompl{m_0/\mpl}

\preprint{IFT-07-08}

\begin{document}
\title{Higgs-Boson Mass Limit \\ within the Randall-Sundrum Model}

\author{Bohdan Grzadkowski}
\email{bohdan.grzadkowski@fuw.edu.pl}
\affiliation{Institute of Theoretical Physics,  University of Warsaw,\\
Ho\.za 69, PL-00-681 Warsaw, Poland}

\author{John Gunion}
\email{jfgucd@physics.ucdavis.edu}
\affiliation{Department of Physics, University of California,\\
Davis CA 95616-8677, USA}

\begin{abstract}
 Perturbative unitarity for $\wlwl$ scattering  is discussed within the Randall-Sundrum model.
 It is shown that the exchange of massive 4D Kaluza-Klein gravitons leads to
 amplitudes growing linearly with the CM energy squared. Summing over
 KK gravitons up to a scale $\lbar$ and testing unitarity at
 $\rts=\lbar$, one finds that unitarity is violated for
 $\lbar$ below the 'naive dimensional analysis' scale, $\lnda$.  It is also
 shown that the exchange of gravitons can substantially relax the upper
 limit from unitarity on the mass of the Standard Model Higgs boson ---
 consistency with unitarity for all $\rts$ below $\lbar$
is possible for $\mh$ as large as $ 1.4\tev$, depending on the
curvature of the background metric.
 Observation of the mass and width (or cross
 section) of one or more KK gravitons at the LHC will directly
 determine the curvature and the scale $\lw$ specifying the couplings of
 matter to the KK gravitons. With this information and a
 measurement of the Higgs boson mass it will be possible to determine
 the precise $\rts$ value 
 below which unitarity will remain valid.
\end{abstract}

\maketitle

\section{Introduction}
Even though the Standard Model (SM) of electroweak interactions perfectly describes 
almost all  existing experimental data, nevertheless the model
suffers from certain theoretical drawbacks. The hierarchy
problem is probably the most fundamental of these: namely, quantum
loop corrections in the SM destabilize the weak
energy scale ${\cal O}(1\tev)$ if the theory is assumed to remain
valid to a much higher scale such as the
Planck mass scale ${\cal O}(10^{19}\gev)$.  Therefore, it is 
believed that the SM is only an effective theory embedded in some 
more fundamental high-scale theory that
presumably could contain gravitational interactions. 
Models that involve extra spatial dimensions could provide a solution
to the hierarchy problem in which gravity plays the major role. 
The most attractive proposal was formulated by
Randall and Sundrum~(RS) ~\cite{Randall:1999ee}. They postulate a 5D
universe with two 4D surfaces (``3-branes''). All the SM particles and
forces with the exception of gravity are assumed to be confined to one
of those 3-branes called the visible or TeV brane.  Gravity lives on
the visible brane, on the second brane (the ``hidden brane'') and in
the bulk.  All mass scales in the 5D theory are of order of the Planck
mass.  By placing the SM fields on the visible brane, all the order
Planck mass terms are rescaled by an exponential suppression factor
(the ``warp factor'') $\Omega_0\equiv e^{-m_0 b_0/2}$, which reduces
them down to the weak scale ${\cal O}(1 \tev)$ on the visible brane
without any severe fine tuning. To achieve the necessary suppression,
one needs $m_0 b_0 /2 \sim 35$. This is a great improvement compared
to the original problem of accommodating both the weak and the Planck
scale within a single theory. 

The RS model is specified by the 5-D action:
\bea
S&=&-\int d^4x\, dy \sqrt{-\what g}\left({2\mplv^3\what R }+\Lambda\right)\non\\
&&+\int d^4x\,\sqrt{-g_{\rm hid}}({\cal L}_{\rm hid}-V_{\rm hid})
+\int d^4x\,\sqrt{-g_{\rm vis}}({\cal L}_{\rm vis}-V_{\rm vis})\,,
\eea
where the notation is self-explanatory, see also \cite{Dominici:2002jv} for details.
In order to obtain a consistent solution to Einstein's
equations corresponding to a low-energy effective  4D theory that is flat, certain conditions
must be satisfied:
$V_{\rm hid}=-V_{\rm vis}={24\mplv^3 m_0 }$ and
$\Lambda=-{24\mplv^3 m_0^2}$. Then, the following metric is a solution of Einstein's equations:
\beq
\what g_{\what\mu\what\nu}(x,y)=\left(
\begin{array}{ccc} 
e^{-2m_0 b_0 |y|}\eta_{\mu\nu}&|& 0 \\
\hline 
0 &|& -b_0^2
\end{array}
\right) \,.
\eeq
After an expansion around the background metric we obtain the gravity-matter interactions
\beq
\call_{\rm int}=-{1\over\lw}\sum_{n\neq 0} h_{\mu\nu}^n T^{\mu\nu} -{\phi_0\over\lphi}T_\mu^\mu
\label{int}
\eeq
where $h_{\mu\nu}^n(x)$ are the Kaluza-Klein (KK) modes (with mass
$m_n$) of the graviton field $h_{\mu\nu}(x,y)$, $\phio(x)$ is the
radion field (the scalar quantum degree of freedom associated with
fluctuations of the distance between the branes), $\lw \simeq \sqrt 2
\mpl \Omega_0$, where $\Omega_0 = e^{-m_0 b_0/2}$, and $\lphi = \sqrt
3\,\lw$.  To solve the hierarchy problem, $\lw$ should be of order
$1-10\tev$, or perhaps higher~\cite{Randall:1999ee}.  In
addition to the radion, the model contains a conventional Higgs boson,
$h$.  The RS model solves the hierarchy problem by virtue of the
fact that the 4D electro-weak scale is given in terms of the ${\cal
  O}(\mpl)$ 5D Higgs vev, $\what v$, by: \beq v_0= \Omega_0 \what v=
e^{-m_0 b_0/2} \what v \sim 1\tev \lsp {\rm for}\lsp m_0 b_0 /2 \sim
35\,.  \eeq

However, the RS model is trustworthy in its own right 
only if the 5D curvature $m_0$ is small compared to the 5D Planck
mass, $\mplv$~\cite{Randall:1999ee}.   The $m_0<\mplv$
requirement and the fundamental RS relation $\mpl^2=2\mplv^3/m_0$
imply that $\mompl=2^{-1/2}(m_0/\mplv)^{3/2}$ should be significantly
smaller than 1.  Hereafter, we 
will focus on the range: $10^{-3} \lsim \mompl \lsim 10^{-1}$.

The goal of this analysis is to determine the cutoff (defined
as some maximum energy up to which the 4D RS theory is well behaved)
and to discuss the unitarity limits on the Higgs boson mass taking into
account KK graviton exchange; for a detailed discussion see \cite{GG}.
%
\section{The cutoff}
The cutoff can be
estimated in a number of ways.  One estimate of the maximum allowed energy
scale is that obtained using the 'naive dimensional analysis' (NDA)
approach~\cite{Manohar:1983md}, the associated scale is denoted by
$\lnda$.~\footnote{The 4D condition for the cutoff $\lnda$ (which
  corresponds to the scale at which the theory becomes strongly
  coupled) is $(\lnda/\lw)^2 N/(4\pi)^2\sim 1$, where $N$ is the
  number of KK-gravitons lighter than $\lnda$ (implying that they
  should be included in the low-energy effective theory). For the RS
  model the graviton mass spectrum for large $n$ is $m_n\simeq m_0\pi
  n \Omega_0$, implying $N\sim \lnda/(m_0\pi\Omega_0)$ which leads to
  Eq.~(\ref{nda}).}  One finds
\beq
\lnda= 2^{7/6}\pi (\mompl)^{1/3}\lw\,,
\label{nda}
\eeq
where $\lw$ was defined in Eq.~(\ref{int}); its inverse sets the
strength of the coupling between matter and gravitons.  We emphasize
that $\lnda$ is obtained when the exchange of the whole tower of KK
modes up to $\lnda$ is taken into account. Physically, $\lnda$ is the energy scale at which the theory
starts to become strongly coupled and string/$M$-theoretic excitations
appear from a 4D observer's point of view ~\cite{Randall:1999ee}.
  In this presentation, we show that
unitarity in the $J=0$ partial wave of $\wlwl$ scattering is always
violated in the RS model for energies below the $\lnda$ scale.  We
will define $\lbar$ as the largest $\rts$ value such that if we sum
over graviton resonances with mass below $\lbar$ (but do not include
diagrams containing the Higgs boson or radion of the model) then
$\wlwl$ scattering remains unitary in the $J=0$ partial wave. 
\begin{figure}[h]
 \bce
 \includegraphics[width=8cm,angle=90]{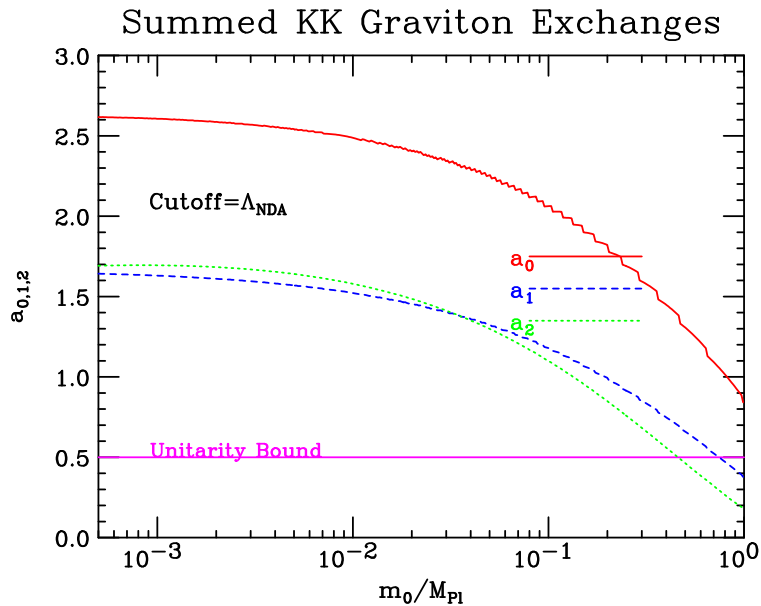}
 \vspace*{-.2cm}
 \caption{We plot $\re a_{0,1,2}$ 
 as  functions of $\mompl$ as computed at $\rts=\lnda$ and summing
 over all KK graviton resonances with mass below $\lnda$, but without
 including Higgs or radion exchanges.
 \label{lndaplot}}
 \ece
\end{figure}
Unitarity of the S-matrix implies that the partial wave amplitudes
$a_J(s)$ must satisfy $|\re a_J|<1/2$. 
As we see from Fig.~\ref{lndaplot}, $\wlwl$
scattering violates unitarity if $\lnda$ is employed as the cutoff.
A more appropriate cutoff is determined 
numerically by requiring $|\re a_{0,1,2}|<1/2$  for $\rts=\lbar$ after
summing over KK resonances with mass below $\lbar$. It is important
to realize that in the presence of KK gravitons and the radion, the
SM cancellation (between Higgs and gauge boson contributions) of 
terms $\propto s$ in the asymptotic behavior of $a_J(s)$ is spoiled, 
that is why graviton contributions turn out to be so relevant.
\begin{figure}[h]
  \bce
  \includegraphics[width=6.5cm,angle=90]{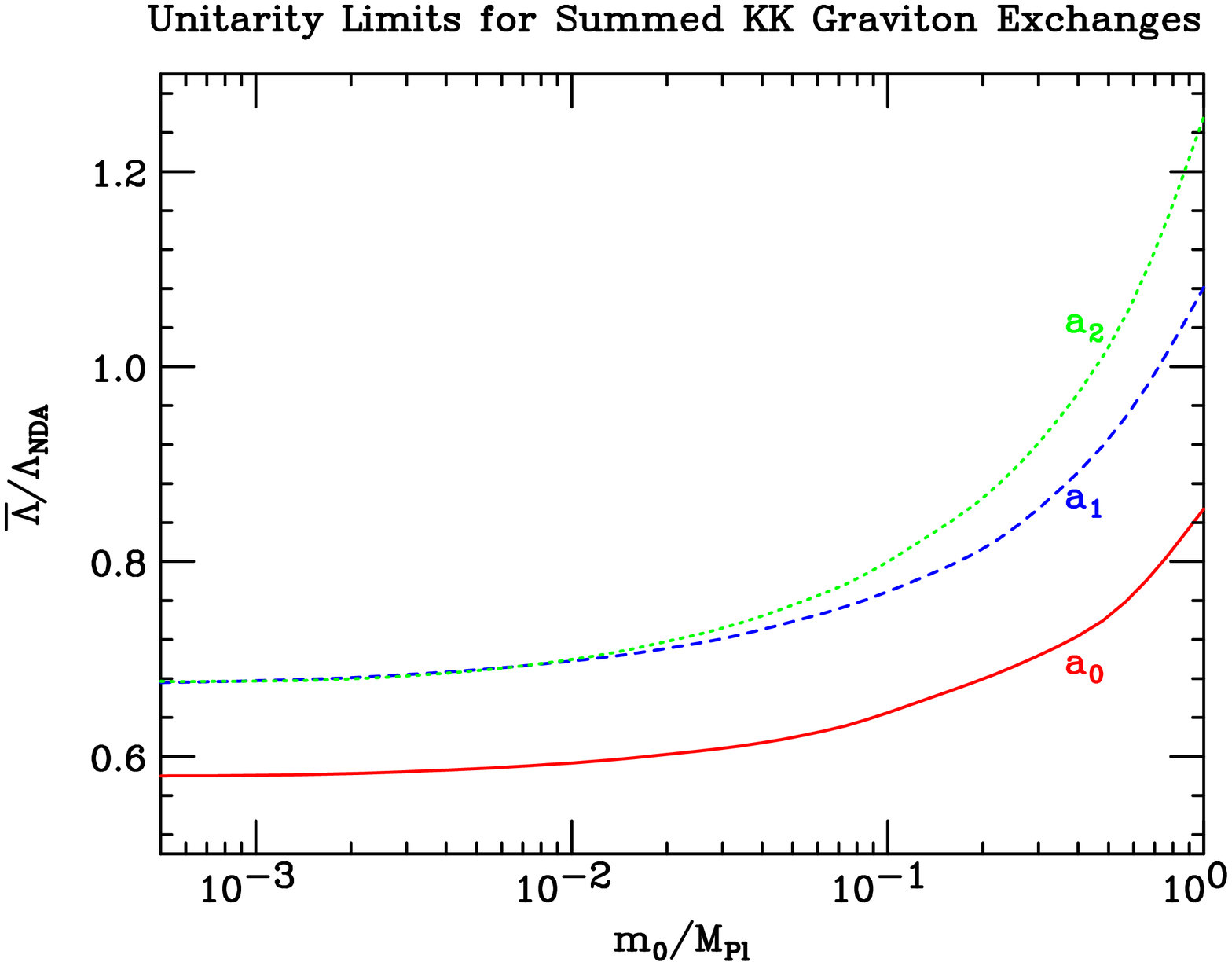}
  \includegraphics[width=6.5cm,angle=90]{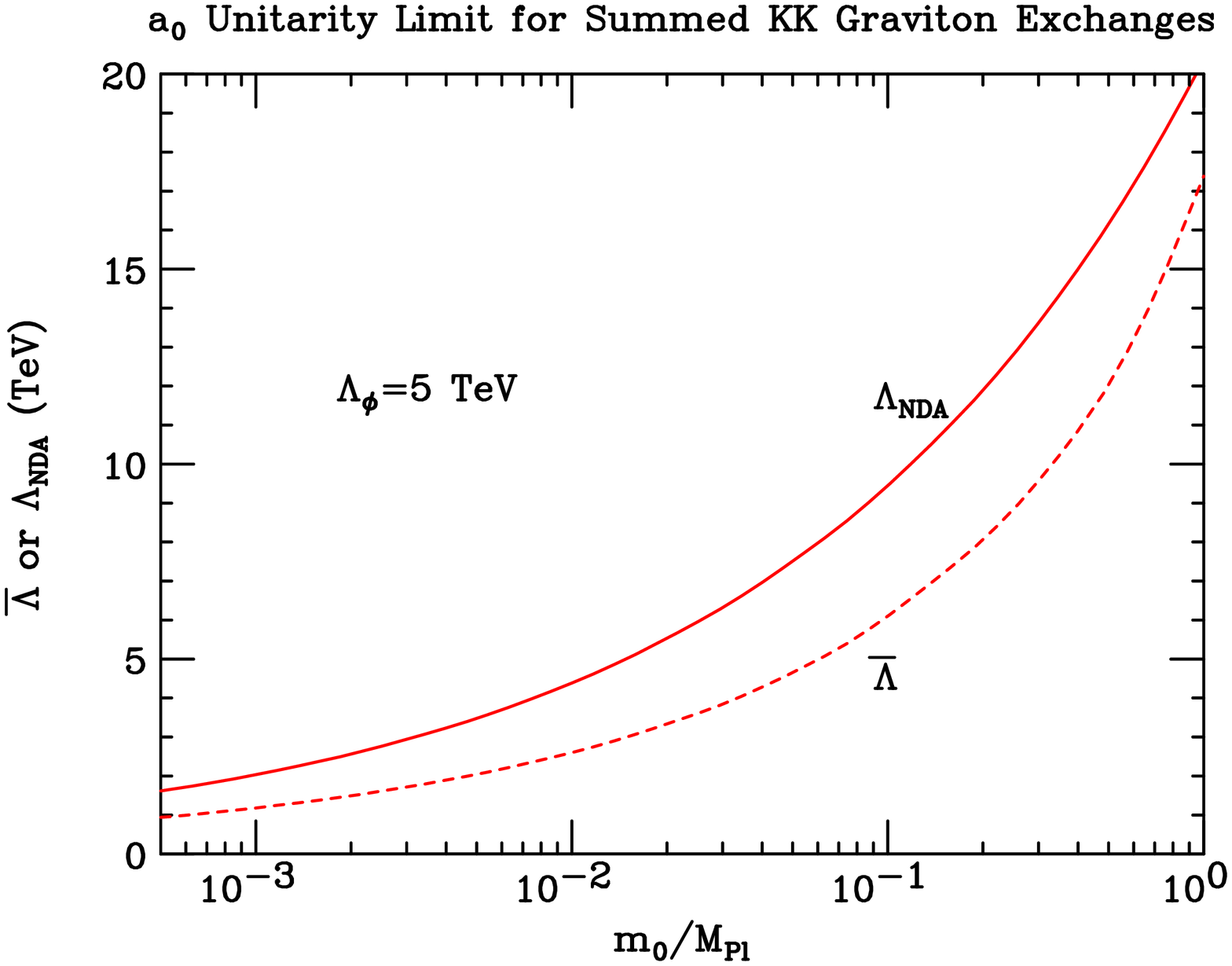}
  \vspace*{-.3cm}
  \caption{In the left hand plot, we give $\lbar/\lnda$ as a function of $\mompl$, where
  $\lbar$ is the largest $\rts$ for which $\wlwl$ scattering is
  unitary after including KK graviton exchanges with mass up to
  $\lbar$, but before including Higgs and radion exchanges. Results
  are shown for the $J=0$, 1 and 2 partial waves. With increasing
  $\rts$ unitarity is always violated earliest in the $J=0$ partial
  wave, implying that $J=0$ yields the lowest $\lbar$. The right hand
  plot shows the individual absolute values of $\lbar(J=0)$ and
  $\lnda$ for the case of $\lphi=5\tev$; $\lbar/\lnda$ is independent
  of $\lphi$
\label{lndaratio}}
\ece
\end{figure}
In the left-hand plot of Fig.~\ref{lndaratio}, we display the ratio
$\lbar/\lnda$ as a function of $\mompl$, where $\lbar$ is the largest
$\rts$ for which $\wlwl$ scattering is unitary when computed including
only the KK graviton exchanges. Results are shown for $J=0$, 1, and 2. As a
function of $\rts$, the $J=0$ partial wave is always the first to
violate unitarity and gives the lowest value of $\lbar$.   We will cut off
our sums over KK exchanges when the KK mass reaches $\lbar$ as
determined by the $J=0$ amplitude.
We see that the $\lbar$ so defined is typically a significant fraction
of $\lnda$, but never as large as $\lnda$.  Still, it is quite
interesting that the unitarity consistency limit $\lbar$ tracks the
'naive' $\lnda$ estimate fairly well as $\mompl$ changes over a wide
range of values (for a qualitative 'derivation' see \cite{GG}).  The
right-hand plot of Fig.~\ref{lndaratio} shows the actual values of
$\lbar$ and $\lnda$ as functions of $\mompl$ for the case of
$\lphi=5\tev$. Note that for larger $\mompl$ they substantially exceed
the input inverse coupling scale $\lphi$, whereas for smaller $\mompl$
they are both substantially below $\lphi$. In other words, using
either $\lbar$ or $\lnda$, one concludes that $\lphi$, and equally
$\lw$, are themselves not appropriate estimators for the maximum
scale of validity of the model.
\begin{figure}[h]
  \bce
  \includegraphics[width=6.5cm,angle=90]{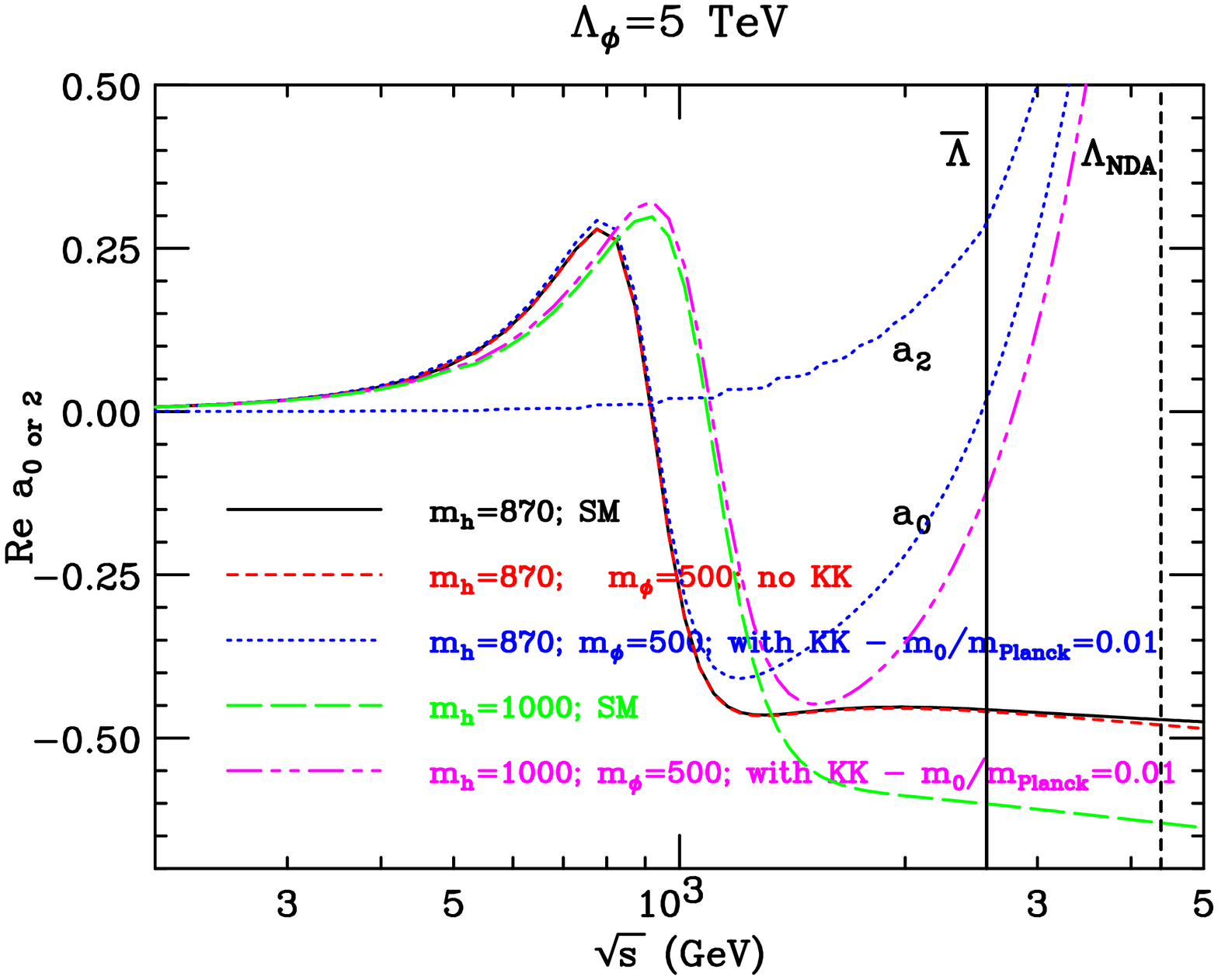}
  \includegraphics[width=6.5cm,angle=90]{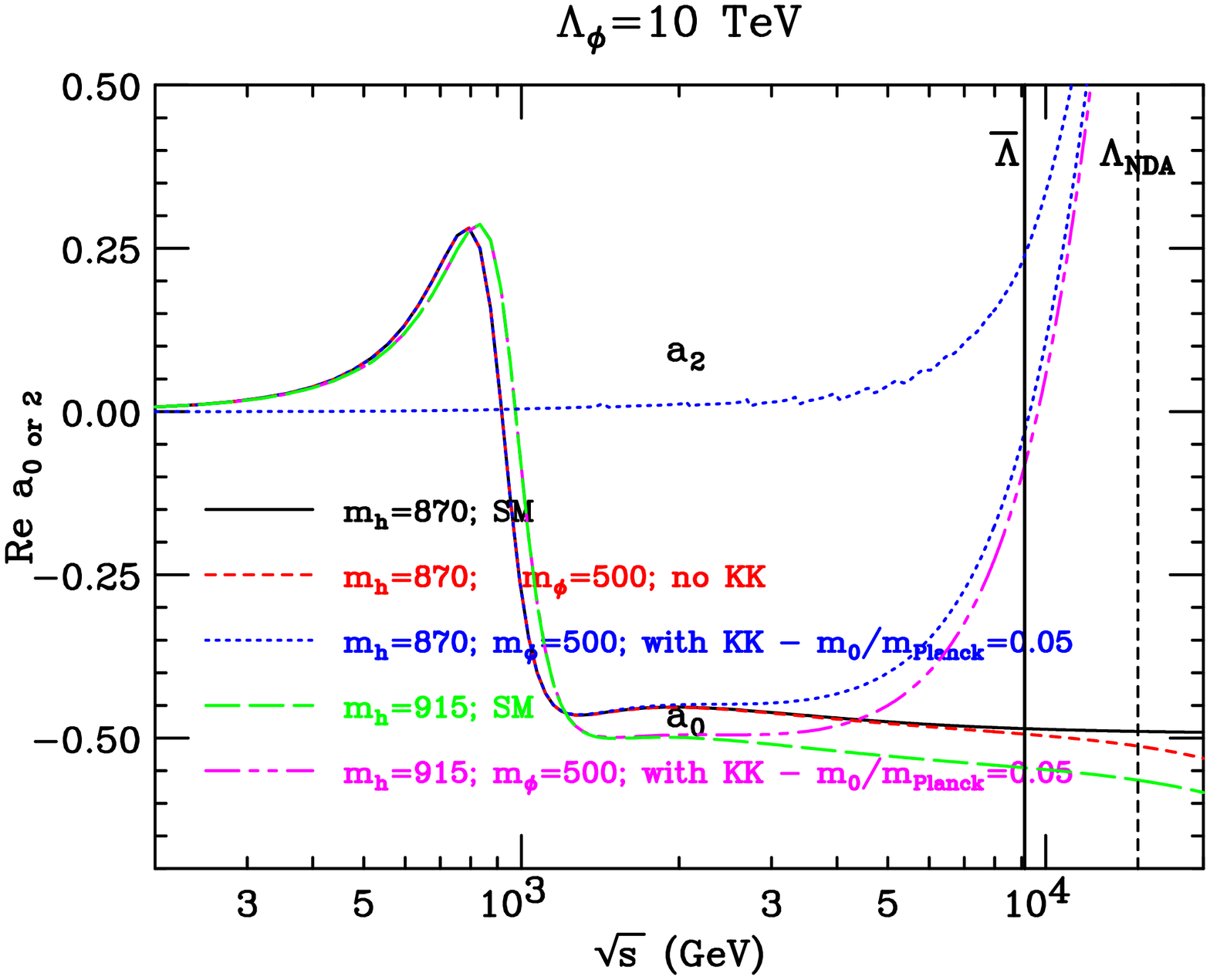}
  \vspace*{-.3cm}
\caption{For $\lphi=5\tev$ --- left ($\lphi=10\tev$ --- right), we plot $\re a_0$ as a function of $\rts$ 
for five cases: 1) solid (black) $\mh=870\gev$, SM contributions only; 2) short dashes (red) $\mh=870\gev$, with an unmixed
 radion of mass $\mphi=500\gev$ included, but no KK gravitons (we do not show
the very narrow $\phi$ resonance);  3) dots (blue) as in 2), but including the sum
over KK gravitons taking $\mompl=0.01$ ($\mompl=0.05$) --- $\re a_2$ is also shown for this case; 4) long
dashes (green) $\mh=1000\gev$ ($915\gev$),
with an unmixed radion of mass $\mphi=500\gev$, but no KK gravitons);
5) as in 4), but including the sum over KK gravitons taking
$\mompl=0.01$ ($\mompl=0.05$). The $\lbar$ and 
$\lnda$ values for $\mompl=0.01$ ($\mompl=0.05$) are indicated by vertical lines. 
\label{svariationcases}}
\ece
\end{figure}
The left-hand plot of Fig.~\ref{svariationcases} shows 
$\re a_0$ as a function of $\rts$
for the case of $\lphi=5\tev$ for two different $\mh$ values and with
and without radion and/or KK gravitons included. In the case where we
include only the SM contributions for $\mh=870\gev$, the figure illustrates unitarity violation as
$\re a_0$ asymptotes to a negative value very close to $-1/2$,
implying that $\mh=870\gev$ is very near the largest value of $\mh$
that is allowed by unitarity in the SM. 
If we add in just the radion contributions (for
$\mphi=500\gev$~\footnote{The radion contribution is always
negligible in the scenario discussed here if $\mphi$ remains
in the range $\mphi\in[10,1000]\gev$ and $\lphi$ is above $1\tev$.
This is however not true in the context of curvature-Higgs mixing 
as discussed in \cite{Grzadkowski:2005tx}.} -- the
$\phi$ resonance is very narrow and is not shown), then a sharp-eyed
reader will see (red dashes) that $\re a_0$ is a bit more negative at
the highest $\rts$ plotted, implying earlier violation of unitarity.
However, if we now include the full set of KK gravitons, which enter
with an increasingly positive contribution, taking $\mompl=0.01$
(dotted blue curve) one is far from violating unitarity due to $\re
a_0<-1/2$ for $\rts$ values above $\mh=870\gev$; instead, the positive
KK graviton contributions, which cure the unitarity problem at
negative $\re a_0$ for $\rts$ above $\mh$, cause unitarity to be
violated at large $\rts$, but {\it above} $\lbar$, as $\re a_0$ passes
through $+1/2$.  In fact, in the case of a heavy Higgs boson we see
that $\re a_2$ actually violates unitarity earlier than does $\re
a_0$. However, even using $\re a_2$ as the criterion, unitarity is
first violated for $\rts$ values above the $\lbar$ value appropriate
to the $\mompl=0.01$ value being considered, but still below $\lnda$.
In fact, it is very generally the case that unitarity is not violated
at $\rts=\lbar$ (which is typically a sizable fraction of $\lnda$) no
matter how small we take $\mompl$. However, as we shall see, unitarity
can be violated in the vicinity of $\rts\sim \mh$ if $\mh$ is large
and $\mompl$ is sufficiently small.

Looking again at the left plot of Fig.~\ref{svariationcases}, we
observe that if $\mh$ is increased to $1000\gev$, the purely SM plus
radion contributions (long green dashes) show strong unitarity
violation at large $\rts$ due to $\re a_0<-1/2$.  However, if we
include the KK gravitons (long dashes and two shorter dashes in
magenta), the negative $\re a_0$ unitarity violation disappears and
unitarity is instead violated at higher $\rts$.  Thus, it is the KK
gravitons that can easily control whether or not unitarity is violated
for $\rts<\lbar$ for a given value of $\mh$.
%
\section{The Higgs-boson mass limit}
\begin{figure}[h]
\bce
\includegraphics[width=6.5cm,angle=90]{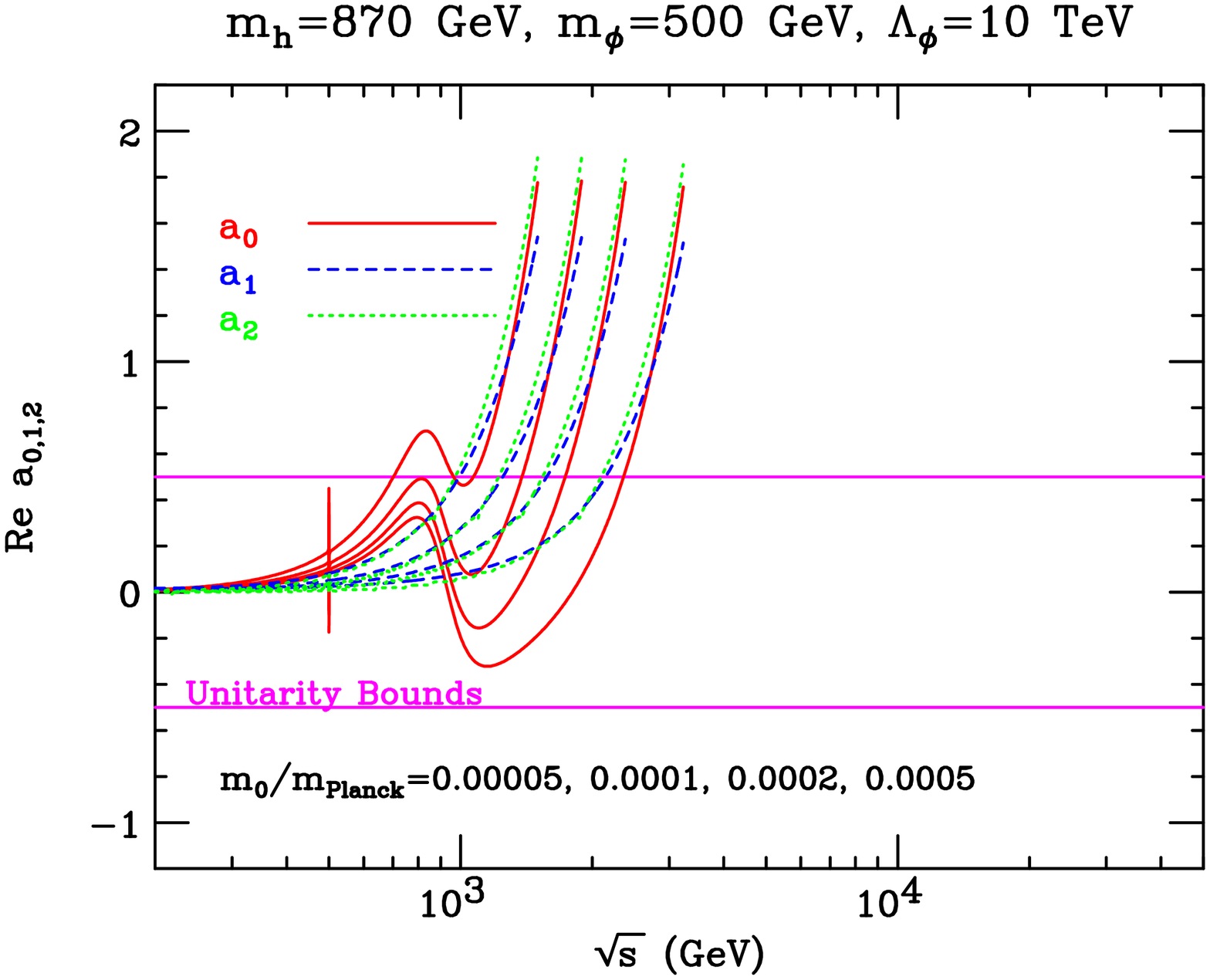}
\includegraphics[width=6.5cm,angle=90]{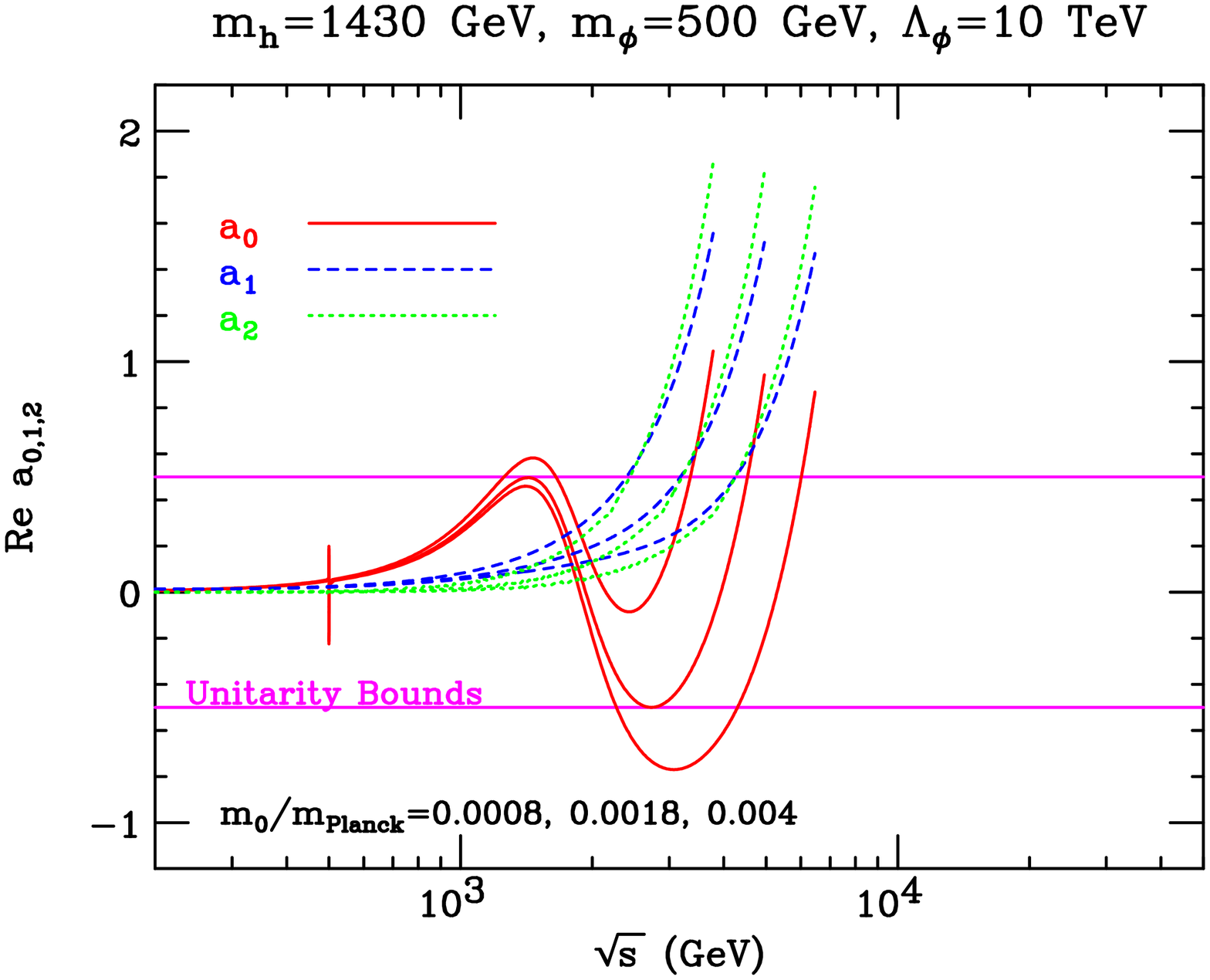}
\vspace*{-.3cm}
\caption{We plot $\re a_{0,1,2}$ as functions of $\rts$ for $\mh=870\gev$ and $\mh=1430\gev$,
  taking
$\mphi=500\gev$ and $\lphi=10\tev$, and for the $\mompl$ values
indicated on the plot. Curves of a given type become higher
as  one moves to lower $\mompl$ values. 
We have included all KK resonances with $m_n<\lbar$ (at
all $\rts$ values). Each curve terminates at $\rts=\lnda$, where
$\lnda$ at a given $\mompl$ is as plotted earlier in
Fig.~\ref{lndaratio}. The value of $\rts$ at which a given curve
crosses above $\re a_0=+1/2$ is always slightly above the $\lbar$
(plotted in Fig.~\ref{lndaratio})
value for the given $\mompl$. \label{a012rts2}}
\ece
\vspace*{-0.8cm}
\end{figure}

As we have already seen, the Higgs plus vector boson exchange contributions have a large
affect on the behavior of $\re a_0$ (whereas the radion exchange
contributions are typically quite small in comparison).  It is
particularly interesting to consider 
cases with a very heavy Higgs boson, focusing on
small values of $\mompl$. 
For $\mh=870\gev$ and $\lphi=10\tev$,  the result appears as the left-hand plot of
Fig.~\ref{a012rts2}.  Note that for the very small value of
$\mompl=0.0001$, unitarity is only just satisfied for $\rts\sim \mh$
and that $\re a_0$ exceeds $+1/2$ near $\rts\sim \mh$.  This is a
general feature in the case of a heavy 
Higgs; there is always a lower bound on $\mompl$ coming purely from
unitarity.  The
right-hand plot of Fig.~\ref{a012rts2} shows how high we can push the
mass of the Higgs boson without violating unitarity. For
$\mh=1430\gev$, we are just barely consistent with the unitarity limit
$|\re a_0|\leq 1/2$ (until large $\rts\gsim \lbar$) if $\mompl=0.0018$
(and $\lphi=10\tev$).  Any lower value of $\mompl$ leads to $\re
a_0>+1/2$ at $\rts\sim\mh$ and any higher value leads to an excursion
to $\re a_0<-1/2$ at higher $\rts$ values (but still below $\lbar$).
There are no experimental limits (coming from
direct production of KK gravitons) of which we are aware
on the $\mompl$ values considered in Fig.~\ref{a012rts2} . For such values, the KK gravitons
would have very small masses, an experimental analysis in that range of $\mompl$ is needed.
\begin{table}[h]
\begin{tabular}{|c|c|c|c|c|}
\hline
$\lphi(\tev)$ &  $5$ & $10$ & $20$ & $40$ \\
\hline\hline
\multicolumn{5}{|c|} {Absolute maximum Higgs mass} \\
\hline
$\mh^{\rm max}(\gev)$ & 1435 & 1430  & 1430 & 1430 \\
required $\mompl$ & $1.32\times 10^{-2}$ & $1.8\times 10^{-3}$ & $2.3\times 10^{-4}$ & $2.9\times
10^{-5}$ \\
associated $m_1(\gev)$ & 103.2 & 28.2 & 7.2 & 1.8 \\ 
\hline
\hline
\multicolumn{5}{|c|}{$\mompl=0.005$: Tevatron limit: $m_1>??$}\\
\hline 
$\mh^{\rm max}(\gev)$ & 1300 & 930 & 920 & 905 \\
associated $m_1(\gev)$ & 39 & 78 & 156 & 313 \\
\hline
\hline
\multicolumn{5}{|c|}{$\mompl=0.01$: Tevatron limit: $m_1>240\gev$}\\
\hline 
$\mh^{\rm max}(\gev)$ & 1405 & 930 & 910 & 895 \\
associated $m_1(\gev)$ & 78 & 156 & 313 & 626 \\
\hline\hline
\multicolumn{5}{|c|}{$\mompl=0.05$: Tevatron limit: $m_1>700\gev$}\\
\hline
$\mh^{\rm max}(\gev)$ & 930 & 915 & 900 & 885 \\
associated $m_1(\gev)$ & 391 & 782 & 1564  & 3129 \\
\hline\hline
\multicolumn{5}{|c|}{$\mompl=0.1$: Tevatron limit: $m_1>865\gev$}\\
\hline
$\mh^{\rm max}(\gev)$ & 920 & 910 & 893 & 883 \\
associated $m_1(\gev)$ & 782 & 1564 & 3128  & 6257 \\
\hline
\end{tabular}
\caption{Unitarity limits on $\mh$ for various $\lphi$ and $\mompl$
values. }
\label{primary}
\end{table}

In Table~\ref{primary}, we summarize the primary implications of our
results by showing a number of limits on $\mh$ for the choices of
$\lphi=5$, 10, 20 and $40\tev$.  The first block gives the very
largest $\mh$ that can be achieved, $\mhmax$, without violating
unitarity in $\wlwl$ scattering for some $\rts<\lbar$, along with the
associated $\mompl$ 
value and mass $m_1$ of the lightest KK graviton.  Unfortunately, no
Tevatron limits (see \cite{magass})
have been given for the associated very small $\mompl$
values.  Even if they end up being experimentally excluded, it is
still interesting from a theoretical perspective that in the RS model
unitarity can be satisfied for all $\rts$ values below the $\lbar$
cutoff of the theory for a Higgs boson mass substantially higher than
the usual $870\gev$ value applicable in the SM context.  One finds
that $\mhmax$ is typically of order $1.4\tev$ if one chooses the
optimal value for $\mompl$ 
(for $\lphi$ in a reasonable range: $5\tev \leq \lphi \leq 40\tev$). 
It is also noteworthy that the required values of $\mompl$ are quite 
consistent with model expectations.

Table~\ref{primary} also gives the $\mhmax$ value achievable for the
four $\lphi$ cases listed above for various fixed $\mompl$.  Also
given are the associated $m_1$ values and the Tevatron direct
production limit when available.  For some of the cases that are
clearly consistent with Tevatron limits, unitarity is satisfied for
$\mh$ values as high as $\sim 915\gev$.

\section{Conclusions}
We have discussed perturbative unitarity for $\wlwl$ within the
Randall-Sundrum theory with two 3-branes and shown 
that the exchange of massive 4D Kaluza-Klein gravitons leads to
amplitudes growing linearly with the CM energy squared.  We have found
that the gravitational contributions cause a violation of unitarity
for $\rts$ below the natural cutoff of the theory, $\lnda$, as
estimated using naive dimensional analysis.

In practice, to determine the cutoff the two basic RS model parameters $\lw$
and $\mompl$ must be extracted from experiment, as should be possible at
the LHC. If the Higgs
mass has also been measured, then the maximum $\rts$ for which $\wlwl$
scattering obeys unitarity in the RS model can be found from the
results of this paper. The most important result obtained here is the 
determination of the maximal Higgs boson mass allowed by requiring
that $\wlwl$ scattering be consistent with unitarity
for all $\rts$ values below the scale $\lbar$ (defined earlier and
always close to $\lnda$): one finds 
$\mh^{\rm max} \leq 1.4\tev$ --- to achieve the upper limit,  
a particular $\lw$-dependent $\mompl$ value is necessary.   

We should emphasize here that we do not need
to consider the effects of the scalar
field(s) that are responsible for stabilizing the inter-brane
separation at the classical level. These fields are normally chosen to be singlets
under the SM gauge groups (sample models include those of
Refs.~\cite{Goldberger:1999wh,Grzadkowski:2003fx}),
and will thus have no direct couplings to the $\wl\wl$ channel. 
For the purpose of this work, the only effect of the inter-brane 
stabilization is to determine the radion mass.

%
\vskip .1in
\centerline{\bf Acknowledgments}
\vskip .1in
This work is
supported in part by the Ministry of Science and Higher Education
(Poland) in years 2006-8 as research project N202~176~31/3844, by EU Marie Curie Research
Training Network HEPTOOLS, under contract MRTN-CT-2006-035505, by the
U.S. Department of Energy grant No. DE-FG03-91ER40674, and by NSF
International Collaboration Grant No. 0218130.  B.G. acknowledges the
support of the European Community under MTKD-CT-2005-029466 Project.

%

\end{document}